\documentclass[twocolumn,prl,aps,superscriptaddress]{revtex4-1}
\usepackage[latin9]{inputenc}
\usepackage{mathtools}
\usepackage{amsbsy}
\usepackage{amstext}
\usepackage[unicode=true,pdfusetitle,
 bookmarks=false,
 breaklinks=false,pdfborder={0 0 1},backref=false,colorlinks=false]
 {hyperref}
\hypersetup{
 bookmarksnumbered=false,bookmarksopen=false}
\makeatletter

\@ifundefined{textcolor}{}
{%
 \definecolor{BLACK}{gray}{0}
 \definecolor{WHITE}{gray}{1}
 \definecolor{RED}{rgb}{1,0,0}
 \definecolor{GREEN}{rgb}{0,1,0}
 \definecolor{BLUE}{rgb}{0,0,1}
 \definecolor{CYAN}{cmyk}{1,0,0,0}
 \definecolor{MAGENTA}{cmyk}{0,1,0,0}
 \definecolor{YELLOW}{cmyk}{0,0,1,0}
}

\usepackage{amsmath}
\usepackage{graphicx}
\usepackage{amssymb}
\usepackage{txfonts,color}
\usepackage{dsfont}
\makeatother

\begin{document}
\title{Hybrid Microwave Radiation Patterns for High-Fidelity Quantum Gates with Trapped Ions}

\author{I. Arrazola}
\affiliation{Department of Physical Chemistry, University of the Basque Country UPV/EHU, Apartado 644, 48080 Bilbao, Spain}
\author{M. B. Plenio}
\affiliation{Institute of Theoretical Physics and IQST, Albert-Einstein-Allee 11, Universit\"at Ulm, D-89069 Ulm, Germany}
\author{E. Solano}
\affiliation{Department of Physical Chemistry, University of the Basque Country UPV/EHU, Apartado 644, 48080 Bilbao, Spain}
\affiliation{IKERBASQUE,  Basque  Foundation  for  Science,  Maria  Diaz  de  Haro  3,  48013  Bilbao,  Spain}
\affiliation{International Center of Quantum Artificial Intelligence for Science and Technology~(QuArtist)
and Department of Physics, Shanghai University, 200444 Shanghai, China}
\author{J. Casanova}
\affiliation{Department of Physical Chemistry, University of the Basque Country UPV/EHU, Apartado 644, 48080 Bilbao, Spain}
\affiliation{IKERBASQUE, Basque  Foundation  for  Science,  Maria  Diaz  de  Haro  3,  48013  Bilbao, Spain}

\begin{abstract}
We present a method that combines continuous and pulsed microwave radiation patterns to achieve robust interactions among hyperfine trapped ions placed in a magnetic field gradient. More specifically, our scheme displays continuous microwave drivings with modulated phases, phase flips, and $\pi$ pulses. This leads to high-fidelity entangling gates which are resilient against magnetic field fluctuations, changes on the microwave amplitudes, and crosstalk effects. Our protocol runs with arbitrary values of microwave power, which includes the technologically relevant case of low microwave intensities. We demonstrate the performance of our method with detailed numerical simulations that take into account the main sources of decoherence.
\end{abstract}

\maketitle

\section{Introduction}
Quantum processors based on trapped ions are leading candidates to build reliable quantum simulators and computers~\cite{Nielsen,Haffner08,Ladd10,Blatt12,Cirac12}. Soon, these could solve computational problems in a more-efficient manner than classical devices by exploiting the quantum correlations among their atomic constituents. To this end, the systematic generation of single-qubit and two-qubit gates with high fidelity is crucial. The latter has been achieved by using laser light that couples the internal (atomic) and external (vibrational) degrees of freedom of the ions, leading to fast single-qubit and two-qubit gates of high fidelity~\cite{Ballance16,Gaebler16,Schafer18}. Nevertheless, scaling laser-based quantum processors while maintaining high fidelities is a hard technological challenge, since it requires the precise control of multiple laser sources.

An alternative approach to laser-driven systems was proposed by Mintert and Wunderlich~\cite{Mintert01}. This involves the use of microwave (MW) fields together with magnetic field gradients to create interactions among the internal states of the ions. Unlike lasers, the control of MW sources is comparatively easy, and their introduction in scalable trap designs is less demanding~\cite{Lekitsch17}. In addition, MW-driven quantum gates do not use any optical transition. This avoids spontaneous emission of some atomic states that define the qubit, which is an unavoidable limiting factor for laser-driven quantum gates~\cite{Plenio97}. Since the initial proposal in~\cite{Mintert01}, the use of MW schemes has been pursued in two distinct fashions, using either static magnetic field gradients~\cite{Mintert01,Weidt15,Piltz16,Welzel19}, or oscillating fields~\cite{Ospelkaus08,Ospelkaus11,Hanh19,Zarantonello19}. In the former, a static magnetic field gradient is used to couple the atomic and vibrational degrees of freedom, while a MW field in the far-field regime is used to modulate this interaction. In the latter, an oscillating MW field in the near-field regime is used as the generator of the qubit-boson interaction. In addition, a method that combines oscillating magnetic field gradients with MW fields on the far-field regime was recently proposed to couple ionic internal and external degrees of freedom~\cite{Srinivas19,Sutherland19}.

Typically, ion qubits are encoded in hyperfine atomic states that are sensitive to magnetic field fluctuations. These are the main source of decoherence, and have to be removed to achieve quantum-information processing with high fidelity. To this end, pulsed and continuous dynamical decoupling (DD) techniques have been introduced~\cite{JonathanPK00,Szwer11,Piltz13,Casanova15,Puebla16,Puebla17,Arrazola18,Wang19}. In particular, the
creation of dressed states has proved useful~\cite{Timoney11,Bermudez12,Cohen15,Wolk17,Webb18} and has led to the highest reported gate fidelities (more than $98\%$) with MW fields in the far-field regime~\cite{Weidt16}. On the other hand, the best near-field MW gates use a driving field on the carrier transition to dynamically decouple the qubits from fluctuations of the qubit frequency~\cite{Harty16}.

In this article, we propose a method to generate two-qubit gates among trapped ions that combines pulsed and continuous MW radiation patterns in the far-field regime. This leads to a scheme that is protected against magnetic fluctuations, errors on the delivered MW fields, and  crosstalk effects caused by the use of long-wavelength MW radiation. Our scheme is flexible since it runs with arbitrary values of the MW power. This includes the relevant case of low-power MWs, which minimizes crosstalk. In particular, inspired by results in Refs.~\cite{CasanovaWS+18, CasanovaWS+19}, our method involves phase-modulated drivings, phase flips, and refocusing $\pi$ pulses, leading to high-fidelity entangling gates within current experimental limitations. We numerically test the performance of our gates in the presence of magnetic fluctuations of different intensities or deviations on the MW Rabi frequencies, as well as under motional heating. We demonstrate the achievement of fidelities largely exceeding $99\%$ in realistic experimental scenarios, while values larger than $99.9\%$ are reachable with small improvements.

\section{The method} 

We consider two $^{171}$Yb$^+$ ions sitting next to each other in the longitudinal direction $(\hat{z})$ of a linear harmonic trap. We define a qubit using two states of the $^2S_{1/2}$  hyperfine manifold. These are $|{\rm g}\rangle\equiv\{F=0,m_F=0\}$ and $|{\rm e}\rangle\equiv\{F=1,m_F=1\}$. Because of the Zeeman effect, the frequency of the $j$th qubit $\omega_j=\omega_0 + \gamma_e B(z^0_j)/2$, where $\omega_0=(2\pi)\times12.6$ GHz, $\gamma_e=(2\pi)\times 2.8$ MHz/G (see Ref.~\cite{Olmschenk07}), and $z^0_j$ is the equilibrium position of the ion. The presence of a constant magnetic field gradient $\partial B/\partial z=g_B$ in the $\hat z$ direction results in different values of $\omega_j$ for each qubit, which allows individual control of each ion with MW fields~\cite{Arrazola18, Piltz14}. The Hamiltonian of the system can be written as
\begin{equation}\label{Hsys}
H = \frac{\omega_1}{2}\sigma_1^z +\frac{\omega_2}{2}\sigma_2^z +\nu a^\dagger a + \eta \nu (a+a^\dagger)S_z \, ,
\end{equation}
where $S_z=\sigma^z_1+\sigma_2^z$,  $a^\dagger$($a$) is the creation (annihilation) operator that corresponds to the center-of-mass mode, $\nu$ is the trap frequency, $\eta=\frac{\gamma_eg_B}{8\nu}\sqrt{\frac{\hbar}{M\nu}}$ is the Lamb-Dicke parameter, which quantifies the strength of the qubit-boson interaction, and $M$ is the mass of a single ion.

Single MW drivings, at detuning $\delta$, can be applied to both ions. Then, Hamiltonian~(\ref{Hsys}) in a rotating frame with respect to $H_0=\frac{\omega_1}{2}\sigma_1^z +\frac{\omega_2}{2}\sigma_2^z +\nu a^\dagger a$ reads (see Appendix A for additional details of the interaction pictures involved)
\begin{equation}\label{HMS}
H= \eta \nu (ae^{-i\nu t}+a^\dagger e^{i\nu t})S_z + \Omega\cos{(\delta t)}S_x.
\end{equation}
For clarity, we have omitted the presence of the breathing mode in Eqs.~(\ref{Hsys}) and~(\ref{HMS}), as well as the crosstalk terms in  Eq.~(\ref{HMS}). However, these will be included in our numerical simulations to demonstrate that they have a negligible impact in our scheme. Furthermore, in Appendix B one can find a complete description of the system Hamiltonian. Now, we move to a second rotating frame with respect to $\Omega\cos{(\delta t)}S_x$ (this is known as the bichromatic interaction picture~\cite{Sutherland19,Roos08}) and use the Jacobi-Anger expansion ($e^{iz\sin{(\theta)}} =  \sum_{n=-\infty}^{+\infty} J_n(z) \ e^{i n\theta}$, and $J_n(z)$ being Bessel functions of the first kind) to obtain
\begin{equation}\label{HMSI}
H= \eta \nu (ae^{-i\nu t}+{\rm H.c.})\Big\{J_0\Big(\frac{2\Omega}{\delta}\Big)S_z+2J_1\Big(\frac{2\Omega}{\delta}\Big)\sin{(\delta t)}S_y\Big\}.
\end{equation}
We keep terms only up to the first order of the Jacobi-Anger expansion, since higher-order terms would not lead to any significant contribution. If we choose $\delta=\nu+\xi$ with $\xi\ll \nu$, and ignore all terms that rotate with $\nu$ by invoking the rotating-wave approximation, we find the gate Hamiltonian
\begin{equation}\label{HMSII}
H_{\rm G} \approx  i\frac{\eta\nu\Omega}{\delta} \Big\{a^\dagger e^{-i\xi t} -{\rm H.c.}\Big\} S_y,
\end{equation}
where we used $J_1(x)\approx x/2$ for small $x$. For evolution times $t_n=2\pi n/\xi$, where $n \in \mathbb{N} $, the time-evolution operator associated with Eq.~$(\ref{HMSII})$ is
\begin{equation}\label{HMSUnitary}
U_{\rm G}(t_n)= \exp{(i\theta_n S_y^2)}
\end{equation}
where $\theta_n=2\pi n \eta^2 \nu^2J_1(2\Omega/\delta)/\xi^2\approx 2\pi n \eta^2 \Omega^2/\xi^2$~\cite{Sorensen99,Sorensen00,Solano99}. By tuning the parameters such that $\theta_n=\pi/8$, the propagator $U_{\rm G}$ evolves the initial (separable) state $|\rm g,g\rangle$ into the maximally entangled Bell state $\frac{1}{\sqrt{2}} (|{\rm g,g}\rangle+i|{\rm e,e}\rangle)$.

To protect this gate scheme from magnetic field fluctuations of the kind $\frac{\epsilon_1(t)}{2}\sigma_1^z+\frac{\epsilon_2(t)}{2}\sigma_2^z$ ($\epsilon_{1,2}(t)$ being stochastic functions), we introduce an additional MW driving that will remove them. We select a MW driving such that  enters in Eq.~(\ref{HMS}) as a carrier term of the form $\frac{\Omega_{\rm DD}}{2}S_y$, leading to
 \begin{equation}\label{HMSDD}
H= \eta \nu (ae^{-i\nu t}+a^\dagger e^{i\nu t})S_z + \Omega\cos{(\delta t)}S_x + \frac{\Omega_{\rm DD}}{2}S_y.
\end{equation}

In the bichromatic picture, Eq.~(\ref{HMSDD}) reads (note that in the following, we adopt the convention $J_{0,1}(2\Omega/\delta) \equiv J_{0,1}$)
\begin{eqnarray}\label{HMSIDD}
H= \eta \nu (ae^{-i\nu t}&+&{\rm H.c.})\Big\{J_0S_z+2J_1\sin{(\delta t)}S_y\Big\} \nonumber \\
&+&\frac{\Omega_{\rm DD}}{2}\Big\{J_0S_y-2J_1\sin{(\delta t)}S_z\Big\}.
\end{eqnarray}
The new driving $\frac{\Omega_{\rm DD}}{2}S_y$ leads to the appearance of the second line in Eq.~(\ref{HMSIDD}). 
Here the $\frac{\Omega_{\rm DD}}{2}J_0 S_y$ term is the responsible for removing magnetic field fluctuations, while $J_1\Omega_{\rm DD}\sin{(\delta t)}S_z$ interferes with the gate and has to be eliminated. This term can ignored under a rotating-wave approximation only if $\Omega_{\rm DD}\ll\delta$, and thus its presence limits the range of applicability of our method since larger 
values of $\Omega_{\rm DD}$ are desirable to better remove the effect of magnetic field fluctuations. To overcome 
this problem, we introduce in all MW drivings [i.e. those leading to the terms $\Omega\cos{(\delta t)}S_x$ and $\frac{\Omega_{\rm DD}}{2}S_y$ in Eq.~(\ref{HMSDD})] a time-dependent phase that will eliminate $J_1\Omega_{\rm DD}\sin{(\delta t)}S_z$. 
This time-dependent phase is
\begin{equation}\label{TP}
\phi(t)=4\frac{\Omega_{\rm DD} J_1}{\delta J_0}\sin^2{(\delta t/2)}.
\end{equation}
\begin{figure}[b]
\centering
\includegraphics[width=1\linewidth]{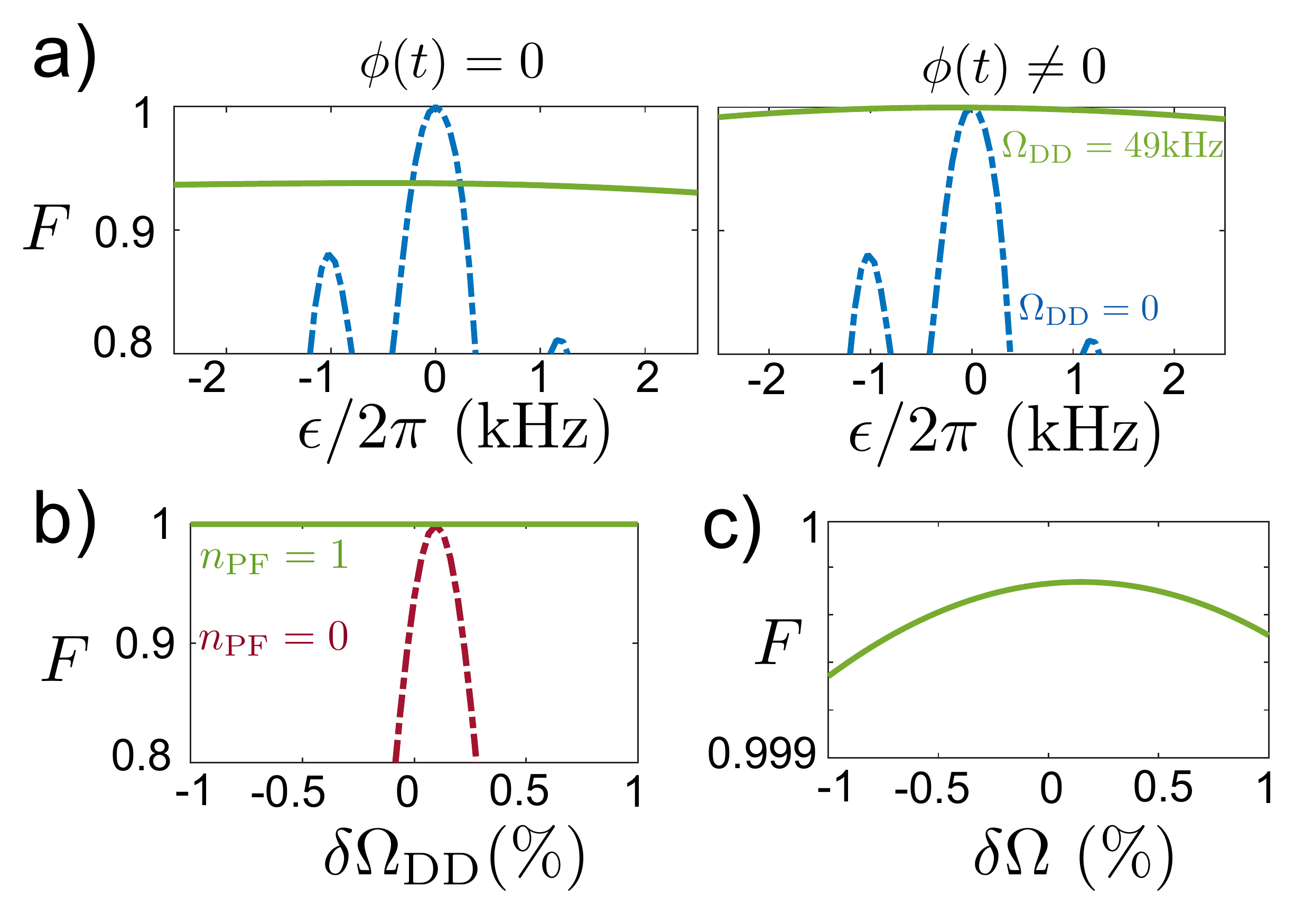}
\caption{Resilience to constant errors for $g_{B}=20.9$ T/m and $\nu=(2\pi)\times138$~kHz ($\eta=0.011$) and $\Omega=(2\pi)\times26$~kHz. (a) Bell state fidelity for $\Omega_{\rm DD}=0$  (blue dashed curve) and $\Omega_{\rm DD}=(2\pi)\times49$~kHz (green solid curve). Right and left panels show the cases with and without phase modulation respectively. (b) Bell state fidelity  for $n_{\rm PF}=1$ (solid green curve) and for $n_{\rm PF}=0$  (red dashed curve). (c) Bell state fidelity with respect to constant shifts in $\Omega(t)$.}\label{Fig1}
\end{figure}
The presence of $\phi(t)$ changes Hamiltonian~(\ref{HMSDD}) to (see Appendix B)
\begin{equation}\label{HMSDDTP}
H= \eta \nu (ae^{-i\nu t}+a^\dagger e^{i\nu t})S_z + \Omega\cos{(\delta t)}S^\parallel_\phi + \frac{\Omega_{\rm DD}}{2}S^{\bot}_\phi,
\end{equation}
where $S^\parallel_\phi \equiv S^+e^{i\phi(t)}+{\rm H.c.}$ and $S^\bot_\phi \equiv -iS^+e^{i\phi(t)}+{\rm H.c.}$ In a rotating frame with respect to $-[\dot{\phi}(t)/2]S_z$ we find
\begin{equation}\label{HMSDDTPbt}
H=\Big\{ \eta \nu (ae^{-i\nu t}+{\rm H.c.})+\frac{\dot{\phi}(t)}{2}\Big\}S_z + \Omega\cos{(\delta t)}S_x+ \frac{\Omega_{\rm DD}}{2}S_y.
\end{equation}
In the bichromatic interaction picture, the previous Hamiltonian transforms as
\begin{eqnarray}\label{HMSIDDI}
\tilde{H}&=&\eta\nu (ae^{-i\nu t}+{\rm H.c.})\Big\{J_0S_z+2J_1\sin{(\delta t)}S_y\Big\} \\
&+&\frac{\tilde{\Omega}_{\rm DD}}{2}S_y -\Omega_{\rm DD}J_1^2/J_0\cos{(2\delta t)}S_y, \nonumber
\end{eqnarray}
where $\tilde{\Omega}_{\rm DD}=J_0\Omega_{\rm DD}(1+2J_1^2/J_0^2)$. Here we can see that, because of the action of $\phi(t)$, the interfering  $J_1\Omega_{\rm DD}\sin{(\delta t)}S_z$ term is removed. Instead, in Eq.~(\ref{HMSIDDI}) we find the term $\Omega_{\rm DD}J_1^2/J_0\cos{(2\delta t)}S_y$, which has a small coupling constant ($\Omega_{\rm DD}J_1^2/J_0$) and that commutes with the gate Hamiltonian~(\ref{HMSII}). In Fig.~\ref{Fig1}(a) (left panel) we show the Bell-state fidelity obtained without phase modulation [i.e. by use of Hamiltonian~(\ref{HMSDD})] for different values of a constant energy deviation $\epsilon$ in the qubit resonance frequencies. The dashed blue curve corresponds to the case $\Omega_{\rm DD}=0$ (i.e. the scheme does not offer protection against $\epsilon$) while the solid green curve incorporates the driving leading to the carrier $\frac{\Omega_{\rm DD}}{2}S_y$. Fig.~\ref{Fig1}(a) (right panel) shows the case with  phase modulation in Eq.~(\ref{HMSDDTP}) that achieves larger fidelities.

\begin{figure}[t]
\centering
\includegraphics[width=1\linewidth]{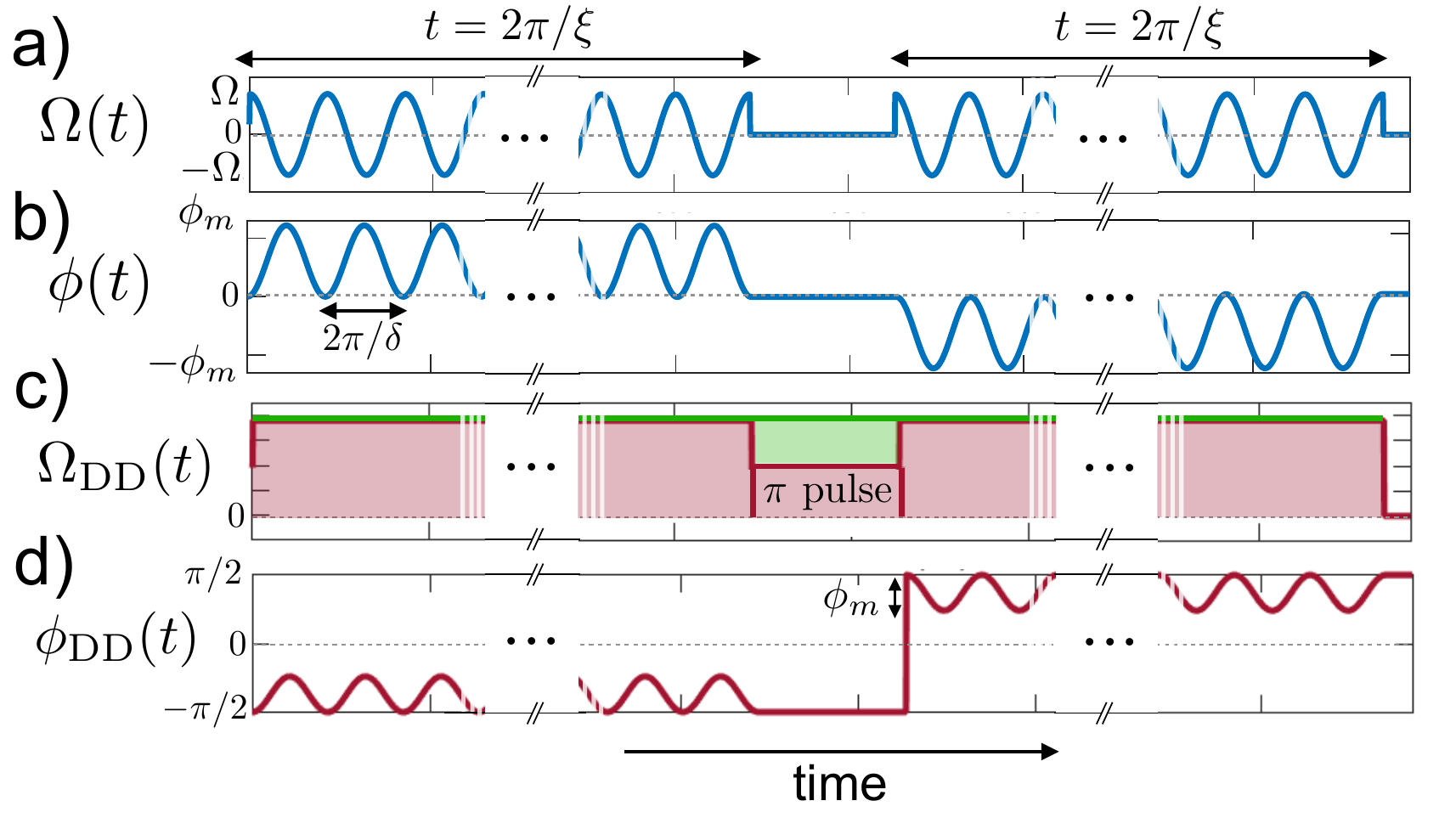}
\caption{Scheme for the control parameters. (a) Bichromatic driving $\Omega(t)=\Omega\cos{(\delta t)}$. (b) Phase modulation of the bichromatic driving. The change of sign during the second half of the evolution is due to the phase flip of the carrier driving. Here $\phi_m=\max{(|\phi(t)|)}$. (c)  Carrier driving acting equivalently on both ions except in the middle, where each ion undergoes a $180^{\circ}$ (red) and a $360^{\circ}$ (green) rotation, respectively. (d) Phase modulation of the DD field. During the second part of the evolution, the phase is flipped from $-\pi/2$ to $\pi/2$. } \label{Fig2}
\end{figure}

Our method can be further improved since the first term in Eq.~(\ref{HMSIDDI}), that is, $\eta\nu (ae^{-i\nu t}+{\rm H.c.}) J_0S_z$, leads to undesired accumulative effects that we can correct. The latter can be calculated in a rotating frame with respect to $\tilde{\Omega}_{\rm DD}S_y/2$ and by computing the second-order Hamiltonian, which reads
\begin{equation}\label{HMSDDIIPrecise}
\tilde{H}\approx H_{\rm G} - g_{\tilde{\Omega}}(2a^\dagger a +1)S_y - \frac{g_{\nu}}{2}(S_x^2+S_z^2) ,
\end{equation}
where $g_{\tilde{\Omega}}=\frac{\tilde{\Omega}_{\rm DD}\eta^2J_0^2}{1-{\tilde{\Omega}^2}_{\rm DD}/\nu^2}$ and $g_\nu=\frac{\nu\eta^2J_0^2}{1-{\tilde{\Omega}^2}_{\rm DD}/\nu^2}$ (see Appendix C). Although small, the terms $ g_{\tilde{\Omega}}(2a^\dagger a +1)S_y $ and $ \frac{g_{\nu}}{2}(S_x^2+S_z^2)$ spoil a superior gate performance. Hence, we eliminate them by introducing refocusing techniques. In particular, to nearly remove the $g_{\tilde{\Omega}}(2a^\dagger a +1)S_y$ term, we divide the evolution into two parts, and flip the phase of the carrier driving during the second part of the evolution. In this respect, the scheme for the control parameters is shown in Fig.~\ref{Fig2}. This phase flip causes a change in the sign of $\Omega_{\rm DD}$, (i.e. $\Omega_{\rm DD} \rightarrow -\Omega_{\rm DD}$) which acts as a refocusing of unwanted shifts in $S_y$. This strategy is also valid to minimize the errors due to constant shifts in $\Omega_{\rm DD}$ as can be seen in Fig.~\ref{Fig1}(b). The phase flip of the carrier forces us to also change the change the sign of $\phi(t)$, since Eq.~(\ref{TP}) should hold during the implementation of the gate. As we see later, performing a large number of phase flips ($n_{\rm PF}$) will further suppress fluctuations on the carrier driving, while it also limits the possible values of $\Omega_{\rm DD}$; see Appendix A for additional details. 

A partial refocusing of the term $\frac{g_{\nu}}{2}(S_x^2+S_z^2)$ in Eq.~(\ref{HMSDDIIPrecise}) is also possible by rotating one of the qubits in the middle and at the end of the gate via $\pi$ pulses. In particular, if these $\pi$ pulses are performed along the $y$ axis, that is, each $\pi$ pulse equals $\exp{(i\pi/2\sigma_1^y)}$, the $S_x^2$ and $S_z^2$ operators change their sign simultaneously, while $S_y^2$ remains unchanged. The combined action of phase flips and $\pi$ pulses allows us to approximate Eq.~(\ref{HMSDDIIPrecise}) as $\tilde{H}\approx H_{\rm G}$. It is important to mention that off-resonant vibrational modes would contribute with accumulative factors similar to the last term in Eq.~(\ref{HMSDDIIPrecise}). These are then refocused by the two $\pi$ pulses as we show in our numerical simulations. As our method removes undesired effects due to additional vibrational modes, it is directly applicable to produce entangling gates between any two ions in a large chain. In addition, one can always consider concatenating sequences of two-qubit operations, leading to multi-qubit gates~\cite{Nielsen}.
For completeness of our analysis, in Fig.~\ref{Fig1}(c) we plot the Bell-state fidelity with respect to constant shifts in $\Omega(t)$. Our scheme also shows robustness with respect to errors of this kind since a shift in  $\Omega(t)$ rotates with frequency $\delta$, which diminishes its effect.

\section{Results} 
\begin{figure}
\centering
\includegraphics[width=1\linewidth]{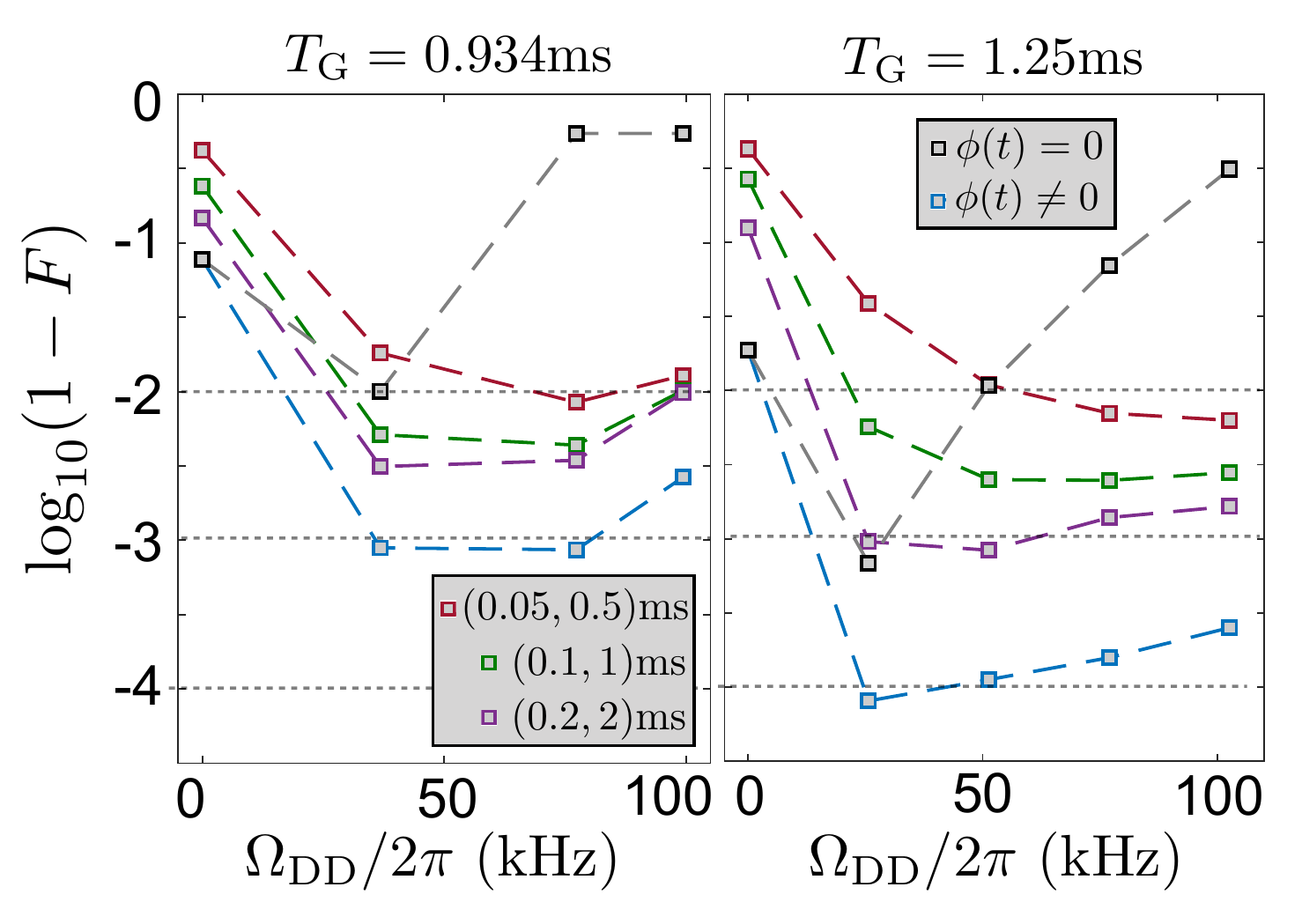}
\caption{Logarithm of the Bell state infidelity as a function of $\Omega_{\rm DD}$, for $g_{B}=20.9$ T/m, $\nu=(2\pi)\times138$~kHz and $\Omega= (2\pi)\times37$~kHz  (left panel), and  $g_{B}=38.5$ T/m, $\nu=(2\pi)\times207$~kHz and $\Omega= (2\pi)\times26.6$~kHz  (right panel). Blue and black squares take into account crosstalk effects and the presence of the off-resonant vibrational mode, with and without phase modulation, respectively. Other curves include motional heating of the center-of-mass mode, and fluctuations of the magnetic field as well as the driving fields. The red, green and purple squares correspond to different error parameters $(\tau,T_2)=(0.05,0.5)$, $(0.1,1)$, and $(0.2,2)$ ms that characterise magnetic field fluctuations.}\label{Fig3}
\end{figure}

To demonstrate the performance of our method in realistic experimental scenarios, we calculate the Bell-state fidelity with fluctuating errors in both magnetic and driving fields, as well as in the presence of motional heating. Furthermore, our simulations include crosstalk terms and an off-resonant motional mode (the initial state of both motional modes is a thermal state with average number of phonons $\bar{n}=1$). The results are shown in Fig.~\ref{Fig3} for two different parameter regimes. These are $g_{B}=20.9$ T/m, $\nu=(2\pi)\times138$~kHz and $\Omega= (2\pi)\times37$~kHz in the left panel and $g_{B}=38.5$ T/m, $\nu=(2\pi)\times207$~kHz and $\Omega= (2\pi)\times26.6$~kHz in the right panel (note both regimes have a Lamb-Dicke parameter $\eta=0.011$).

Blue squares indicate Bell-state infidelities obtained with our method. Thirty-one phase flips are used. For ${\Omega}_{\rm DD}=0$, the fidelities are below $99\%$ even without fluctuating errors. We identify that this is due to the crosstalk of the MW driving fields at Rabi frequency $\Omega$, since these induce frequency shifts in the off-resonant qubits. If $\Omega_{\rm DD} \neq 0$, these energy shifts are canceled and we find fidelities ranging from $99.9\%$ to $99.99\%$. These values are obtained with moderate MW radiation power rather than with Rabi frequencies on the order of megahertz; for the latter, see Ref.~\cite{Arrazola18}. The parameters in the right panel in Fig.~\ref{Fig3} are more favourable for several reasons: First, the Rabi frequency is smaller than for the case in the left panel and the magnetic field gradient is larger; both lower crosstalk effects. Second, a higher trap frequency lowers the effect of the off-resonant mode. Finally, a smaller $\Omega/\delta$ ratio is also preferable to avoid any effect of higher-order Bessel functions. In this respect, note we always truncate the Jacobi-Anger expansion to the first order; see Eq.~(\ref{HMSIDDI}). Black squares indicate the infidelities obtained without phase modulation. As expected, phase modulation is crucial to remove energy shifts induced by ${\Omega}_{\rm DD}$.
Other curves take into account the heating of the center-of-mass mode and fluctuating errors in the magnetic field and MW drivings. The effect of the former is introduced in our model with a dissipative term of the form
$\frac{\Gamma}{2}\big\{(\bar{N}+1)(2a\rho a^\dagger - a^\dagger a \rho - \rho a^\dagger a) + \bar{N}(2a^\dagger\rho a - a a^\dagger \rho - \rho a a^\dagger)\big\}$,
where $\dot{\bar{n}}=\Gamma\bar{N}$ is the heating rate and $\bar{N}=1/(e^{\hbar\nu/k_{\rm B}T}-1)$, where $T=300$K. In the left panel in Fig.~\ref{Fig3}, we consider a heating rate of $\dot{\bar{n}}\approx300$ phonons/s~\cite{Weidt15,Weidt16,Brownnutt15}. For the right panel in Fig.~\ref{Fig3}, we consider a more-favourable scenario with $\dot{\bar{n}}\approx200$ phonons/s. 

Magnetic and MW fluctuations are introduced via an Ornstein-Uhlenbeck stochastic process~\cite{Gillespie96}. Each point in Fig.~\ref{Fig3} corresponds to 100 realizations. The Ornstein-Uhlenbeck process is characterized by the correlation time $\tau_{B}$ and coherence time $T_2$ for the magnetic field fluctuations, while $\tau_\Omega$ and the relative amplitude error $\delta_{\Omega}$ are used for the MW-driving-field fluctuations. For the driving fields, we choose correlation time $\tau_B=500~\mu$s and relative amplitude error $\delta_{\Omega}=0.5\%$ in the left panel, and correlation time $\tau_\Omega=1$~ms and $\delta_{\Omega}=0.25\%$ in the right panel~\cite{Cohen16}. Different strengths for the magnetic field fluctuations are given by the red, green and purple squares, with parameters $(\tau,T_2)=(0.05,0.5)$, $(0.1,1)$, and $(0.2,2)$~ms, respectively.

Our numerical simulations predict fidelities greater $99\%$ even for the worst case corresponding to $(\tau,T_2)=(0.05,0.5)$ ms and with current heating rates (left panel). In a more-optimistic experimental scenario with $(\tau,T_2)=(0.2,2)$ ms, our protocol leads to fidelities larger than $99.9\%$ for distinct values of $\Omega_{\rm DD}$ (right panel). 

\section{Conclusions}
We propose a method for the generation of entangling gates that combines phase-modulated continuous MW drivings with phase flips and refocusing $\pi$ pulses to produce entangling gates with high fidelity. Numerical simulations including the main sources of decoherence show that fidelities on Bell-state preparation exceeding $99\%$ are possible within current experimental limitations, while fidelities larger than $99.9\%$ are achievable with further experimental improvement.

\section*{acknowledgements}
We acknowledge financial support from Spanish Government via PGC2018-095113-B-I00 (MCIU/AEI/FEDER, UE), Basque Government via IT986-16, and QMiCS (820505) and OpenSuperQ (820363) of the EU Flagship on Quantum Technologies, and the EU FET Open Grant Quromorphic. I. A. acknowledges support from Basque Government Ph.D. Grant No. PRE-2015-1-0394. M. B. P. acknowledges support by the ERC Synergy Grant BioQ, the EU Flagship project AsteriQs, and the BMBF projects Nanospin and DiaPol. J. C. acknowledges support from Juan de la Cierva Grant No. IJCI-2016-29681 and the UPV/EHU grant EHUrOPE.

\section*{Appendix A: Interaction pictures and microwave sequence }

Our reference frame is established by $H_0=\frac{\omega_1}{2}\sigma_1^z + \frac{\omega_2}{2}\sigma_2^z +\nu a^\dagger a$, named the ``ion frame''. This means that the Bell state produced by the protocol will be rotating according to this frame. We make use of various interaction pictures to obtain a better understanding of all interactions. The first one is the so-called bichromatic interaction picture~\cite{Sutherland19,Sutherland20,Roos08}, and it is defined by $H_{\rm B}(t)=\Omega\cos{(\delta t)} S_x$. As the gate Hamiltonian in Eq.~(\ref{HMSII}) is written in this picture, the state produced at time $t_n$ will match the one in the ion frame if $U_1(t_n)=\exp{[i\Omega {\rm sinc}{(\delta t_n)}]}=\mathds{1}$. For that, we impose the following relation between the detunings $\xi=\delta/N=\nu/(N-1)$, thus ensuring that $\delta t_n=2\pi N$, where $N \in \mathbb{N}$. The latter is not strictly necessary, as the same effect can be obtained by pulse shaping~\cite{Sutherland19,Roos08}. 

The addition of the carrier driving introduces the term $\frac{\tilde{\Omega}_{\rm DD}}{2}S_y$ in Eq.~(\ref{HMSIDDI}), which,  while commuting with the gate Hamiltonian, produces qubit rotations that have to be synchronized with the time $t_n$. For that, it is necessary to choose $\tilde{\Omega}_{\rm DD}$ in such a way that $\tilde{\Omega}_{\rm DD} t_n=2\pi m$ where $m \in \mathbb{N} $. Moreover, the number of phase flips also limits the possible values of ${\Omega}_{\rm DD}$ as the time interval between phase flips needs to be a multiple of $2\pi/\tilde{\Omega}_{\rm DD}$. The latter is crucial for a correct elimination of magnetic field fluctuations. At the same time, this also sets a lower bound for the minimum nonzero value of ${\Omega}_{\rm DD}$, which increases with the number of phase flips. Finally, in the phase-modulated case, the time interval between two phase flips should also be a multiple of $\pi/\delta$.

\section*{Appendix B: Hamiltonian with crosstalk terms and two vibrational modes }

The Hamiltonian describing two trapped ions under a static magnetic field gradient $\partial B/\partial z=g_B$, in an interaction picture with respect to the free-energy term of the qubits and the modes $\frac{\omega_1}{2}\sigma_1^z+\frac{\omega_2}{2}\sigma_2^z + \nu a^\dagger a + \sqrt{3}\nu b^\dagger b$ is
\begin{eqnarray}\label{Hzero}
H_{sys}&=& \eta \nu (ae^{-i\nu t}+a^\dagger e^{i\nu t})(\sigma_1^z+\sigma_2^z) \nonumber\\ &+& 3^{-1/4}\eta\nu(be^{-i\sqrt{3}\nu t}+b^\dagger e^{i\sqrt{3}\nu t})(-\sigma_1^z+\sigma_2^z)
\end{eqnarray}
where $a^\dagger$($a$) and $b^\dagger$ ($b$) are the creation (annihilation) operators associated to the longitudinal center-of-mass and breathing modes of the two-ion crystal. The former is used as the quantum bus of the qubits, while the latter stays off-resonant. For the numerical simulations in the main text, we truncate the Hilbert space of these at $N_{c.m.}=15$ and $N_{br}=5$ respectively. 

The Hamiltonian of a microwave driving with frequency $\tilde{\omega}_1$ and phase $\phi$ acting on the first qubit reads
\begin{equation}\label{Hq1}
H_{q1}=\frac{\Omega_1}{2}(\sigma_1^+ e^{i\omega_1 t} + \sigma_2^+e^{i\omega_2 t})(e^{i\tilde{\omega}_1 t}e^{-i\phi} +e^{-i\tilde{\omega}_1 t}e^{i\phi}) +{\rm H.c.},
\end{equation}
where $|\omega_1- \tilde{\omega}_1 |\ll \omega_2-\omega_1 \ll \omega_1+\omega_2$. If we ignore terms rotating with $\pm(\omega_1+\tilde{\omega}_1)$ and $\pm(\omega_2+\tilde{\omega}_1)$ by applying the rotating-wave approximation and consider $\tilde{\omega}_1=\omega_1$, the Hamiltonian reads
\begin{equation}\label{Hq1_2}
H_{q1}=\frac{\Omega_1}{2}(\sigma_1^+ + \sigma_2^+e^{i\Delta\omega t})e^{i\phi} +{\rm H.c.},
\end{equation}
where $\Delta\omega=\omega_2-\omega_1=\frac{\gamma_e g_B}{2}\Big(\frac{2e^2}{4\pi\epsilon_0 M \nu^2}\Big)^{1/3}$. In our protocol, each qubit is driven by two phase-modulated microwave fields, and the total Hamiltonian reads 
\begin{eqnarray}\label{Hdriving}
H&=&H_{sys} + \sum_{j=1}^2\Big\{2\Omega\cos{(\delta t)} \cos{[\omega_j t - \phi(t)]} \nonumber \\  &-&  \Omega_\textrm{DD}\sin{[\omega_j t - \phi(t)]}\Big\} (\sigma_1^+ e^{i\omega_1 t} + \sigma_2^+e^{i\omega_2 t} +\textrm{H.c.}).
\end{eqnarray}
After all the terms rotating with frequencies on the order of $\omega_j$ have in ignored, the Hamiltonian is reduced to
\begin{eqnarray}\label{Htotal}
H&=&H_{sys}  +[\Omega(t) -if(t)\frac{\Omega_\textrm{DD}}{2}]e^{if(t)\phi(t)} \nonumber \\ &\times& [\sigma_1^+ (1+e^{-i\Delta\omega t }) + \sigma_2^+(1+e^{i\Delta\omega t })] +\textrm{H.c.},
\end{eqnarray}
where $f(t)$ changes from $+1$ to $-1$ depending on the number of phase flips applied, and $\phi(t)$ can be found in the main text. This is the Hamiltonian used in our numerical simulations, which includes the crosstalk terms that rotate with $\pm\Delta\omega$ as they have a non-negligible effect. However, these will be averaged out by the introduction of the carrier field. For clarity, in the main text we disregard these crosstalk terms and the off-resonant motional mode, which reduces the Hamiltonian (\ref{Htotal}) to
\begin{eqnarray}\label{Htext}
H=H_{sys} + [\Omega\cos{(\delta t)} -if(t)\frac{\Omega_\textrm{DD}}{2}]e^{if(t)\phi(t)}S_+ + \textrm {H.c.},
\end{eqnarray}
which corresponds to Eq.~(\ref{HMSDDTP}) except for the $f(t)$ and the off resonant mode, which are not included in the main text for simplicity.

\section*{Appendix C: Second-Order Hamiltonian derivation in the phase modulated case}

In this section we derive Eq.~(\ref{HMSDDIIPrecise}), starting from Hamiltonian~(\ref{HMSIDDI}). Moving to an interaction with respect to $\tilde{\Omega}_{\rm DD}S_y/2$, Eq.~(\ref{HMSIDDI}) transforms into
\begin{eqnarray}\label{GateHamil2}
\tilde{H} &=&\eta \nu J_0 (\tilde{S}_+a e^{-i(\nu-\tilde{\Omega}) t}+\tilde{S}_-a e^{-i(\nu+\tilde{\Omega}) t} + {\rm H.c.}) \nonumber\\ &+& 2\eta \nu J_1(a e^{-i\nu t} + a^\dagger e^{i\nu t})\sin{(\delta t)}S_y  \nonumber\\ &-& \tilde{\Omega}_{\rm DD}\frac{{J^2_1}}{J_0}\cos{(2\delta t)}S_y, 
\end{eqnarray}
where $\tilde{\Omega}\equiv\tilde{\Omega}_{\rm DD}$ and $\tilde{S}_+=\frac{1}{2}(S_z\pm i S_x)$. 

It is convenient to calculate the second-order Hamiltonian that corresponds to the first term in Eq.~(\ref{GateHamil2}), as its coupling strength is larger than that of the gate, in the regime with $\tilde{\Omega}_{\rm DD} < \nu$. The second-order Hamiltonian can be extracted by use of the Magnus expansion, and it is given by
\begin{eqnarray}\label{SecondOrder1}
\tilde{H}^{(2)}(t)=&-&i\frac{\eta^2\nu^2J_0^2}{2}\int_0^{t} dt' [\tilde{S}_+a e^{-i(\nu-\tilde{\Omega}) t}+\tilde{S}_-a e^{-i(\nu+\tilde{\Omega}) t}\nonumber \\ &+& {\rm H.c.},\tilde{S}_+a e^{-i(\nu-\tilde{\Omega}) t'}+\tilde{S}_-a e^{-i(\nu+\tilde{\Omega}) t'} + {\rm H.c.}].
\end{eqnarray}
First, we can calculate the part that will lead to the time-independent, accumulative part. These terms are
\begin{eqnarray}\label{TimeInde1}
\int_{0}^tdt'[a \tilde{S}_+,a^\dagger \tilde{S}_-]e^{-i(\nu-\tilde{\Omega})(t-t')}+{\rm H.c.}\nonumber\\ = [a \tilde{S}_+,a^\dagger \tilde{S}_-] \frac{-2i}{\nu-\tilde{\Omega}}(1-\cos{[(\nu-\tilde{\Omega} )t]}) \\
\int_{0}^tdt'[a \tilde{S}_-,a^\dagger \tilde{S}_+]e^{-i(\nu+\tilde{\Omega})(t-t')}+{\rm H.c.}\nonumber\\ = [a \tilde{S}_-,a^\dagger \tilde{S}_+] \frac{-2i}{\nu+\tilde{\Omega}}(1-\cos{[(\nu+\tilde{\Omega} )t]}).
\end{eqnarray}
The commutators can be calculated, with use of the relation $[AB,A^\dagger B^\dagger]=[A,A^\dagger]BB^\dagger + A^\dagger A[B,B^\dagger]$, which is valid if $[A,B]=0$. Using this, we have 
\begin{eqnarray}\label{Commutators1}
[a \tilde{S}_+, a^\dagger \tilde{S}_-]&=&[a,a^\dagger]\tilde{S}_+\tilde{S}_- + a^\dagger a[\tilde{S}_+,\tilde{S}_-]\nonumber\\ &=&\frac{1}{4}(S_x^2+S_z^2) + (a^\dagger a +\frac{1}{2})S_y \\ 
 \ [a \tilde{S}_-, a^\dagger \tilde{S}_+]&=&[a,a^\dagger]\tilde{S}_-\tilde{S}_+ + a^\dagger a[\tilde{S}_-,\tilde{S}_+]\nonumber\\ &=&\frac{1}{4}(S_x^2+S_z^2) - (a^\dagger a +\frac{1}{2})S_y,
\end{eqnarray}
and the time-independent part of the second-order Hamiltonian reads
\begin{eqnarray}\label{SecondOrder2}
\tilde{H}_{\rm eff}^{(2)}=&-&\frac{\eta^2\nu^2J_0^2}{4}\Big\{\frac{1}{\nu-\tilde{\Omega}} +\frac{1}{\nu+\tilde{\Omega}}\Big\}(S_x^2+S_z^2) \nonumber\\&-&\frac{\eta^2\nu^2J_0^2}{2}\Big\{\frac{1}{\nu-\tilde{\Omega}} -\frac{1}{\nu+\tilde{\Omega}}\Big\}(2a^\dagger a+1) S_y,
\end{eqnarray}
which can be rewritten as
\begin{equation}\label{SecondOrder3}
\tilde{H}_{\rm eff}^{(2)}=-\frac{g_\nu}{2}(S_x^2+S_z^2) -g_{\tilde{\Omega}}(2a^\dagger a +1)S_y,
\end{equation}
where $g_\nu=\frac{\nu\eta^2J_0^2}{1-\tilde{\Omega}^2/\nu^2}$ and $g_{\tilde{\Omega}}=\frac{\tilde{\Omega}\eta^2J_0^2}{1-\tilde{\Omega}^2/\nu^2}$.


\begin{thebibliography}{99}

\bibitem{Nielsen} M. A. Nielsen and I. L. Chuang, {\it Quantum Computation and Quantum Information} (Cambridge University press, Cambridge, 2000).

\bibitem{Haffner08} H. H\"affner, C. F. Roos, and R. Blatt, Quantum computing with trapped ions, \href{https://www.sciencedirect.com/science/article/abs/pii/S0370157308003463}{Phys. Rep. {\bf469}, 155 (2008).}

\bibitem{Ladd10} T. D. Ladd, F. Jelezko, R. Laflamme, Y. Nakamura, C. Monroe, and J. L. O' Brien, Quantum computers, \href{https://www.nature.com/articles/nature08812}{Nature {\bf464}, 45 (2010).}

\bibitem{Blatt12} R. Blatt and C. F. Roos, Quantum simulations with trapped ions, \href{https://www.nature.com/articles/nphys2252?draft=collection}{Nat. Phys. {\bf8}, 277 (2012).}

\bibitem{Cirac12} J. I. Cirac and P. Zoller, Goals and opportunities in quantum simulation, \href{https://www.nature.com/articles/nphys2275}{Nat. Phys. {\bf8}, 264 (2012).}

\bibitem{Ballance16} C. J. Ballance, T. P. Harty, N. M. Linke, M. A. Sepiol, and D. M. Lucas, High-fidelity quantum logic gates using trapped-ion hyperfine qubits, \href{https://journals.aps.org/prl/abstract/10.1103/PhysRevLett.117.060504}{Phys. Rev. Lett. {\bf 117}, 060504 (2016).}

\bibitem{Gaebler16} J. P. Gaebler, T. R. Tan,  Y. Lin, Y. Wan, R. Bowler, A. C. Keith, S. Glancy, K. Coakley, E. Knill, D. Leibfried, and D. J. Wineland, High-fidelity universal gate set for $^9{\rm Be}^+$ ion qubits, \href{https://journals.aps.org/prl/abstract/10.1103/PhysRevLett.117.060505}{Phys. Rev. Lett. {\bf 117}, 060505 (2016).}



\bibitem{Schafer18} V. M. Sch\"afer, C. J. Ballance, K. Thirumalai, L. J. Stephenson, T. G. Ballance, A. M. Steane, D. M. Lucas, Fast quantum logic gates with trapped-ion qubits, \href{https://www.nature.com/articles/nature25737}{Nature {\bf555}, 75 (2018).}

\bibitem{Mintert01} F. Mintert, and C. Wunderlich, Ion-trap quantum logic using long-wavelength radiation, \href{https://journals.aps.org/prl/abstract/10.1103/PhysRevLett.87.257904}{Phys. Rev. Lett. {\bf 87}, 257904 (2001).}

\bibitem{Lekitsch17} B. Lekitsch, S. Weidt, A. G. Fowler, K. M\o lmer, S. J. Devitt, C. Wunderlich, and W. K. Hensinger, Blueprint for a microwave trapped ion quantum computer, \href{http://advances.sciencemag.org/content/3/2/e1601540}{Sci. Adv. {\bf 3}, e1601540 (2017).}

\bibitem{Plenio97}  M. B. Plenio, and P. L. Knight, Decoherence limits to quantum computation using trapped ions, \href{http://rspa.royalsocietypublishing.org/content/453/1965/2017}{Proc. Roy. Soc. A {\bf 453}, 2017 (1997).}



\bibitem{Weidt15} S. Weidt, J. Randall, S.~C. Webster, E. D. Standing, A. Rodriguez, A. E. Webb, B. Lekitsch, and W. K. Hensinger, Ground-state cooling of a trapped ion using long-wavelength radiation, \href{https://journals.aps.org/prl/abstract/10.1103/PhysRevLett.115.013002}{Phys. Rev. Lett. {\bf 115}, 013002 (2015).}

\bibitem{Piltz16} Ch. Piltz, T. Sriarunothai, S.~S. Ivanov,  S. W\"olk, and C. Wunderlich, Versatile microwave-driven trapped ion spin system for quantum information processing, \href{http://advances.sciencemag.org/content/2/7/e1600093}{Sci. Adv. {\bf 2}, e1600093 (2016).}



\bibitem{Welzel19} J. Welzel, F. Stopp and F. Schmidt-Kaler, Spin and motion dynamics with zigzag ion crystals in transverse magnetic gradients, \href{https://iopscience.iop.org/article/10.1088/1361-6455/aaf347}{J. Phys. B: At. Mol. Opt. Phys. {\bf52}, 025301 (2019).}


\bibitem{Ospelkaus08} C. Ospelkaus, C. E. Langer, J. M. Amini, K. R. Brown, D. Leibfried, and D. J. Wineland, Trapped-ion quantum logic gates based on oscillating magnetic fields, \href{https://journals.aps.org/prl/abstract/10.1103/PhysRevLett.101.090502}{Phys. Rev. Lett. {\bf 101}, 090502 (2008).}

\bibitem{Ospelkaus11} C. Ospelkaus, U. Warring, Y. Colombe, K. R. Brown, J. M. Amini, D. Leibfried, and D. J. Wineland, Microwave quantum logic gates for trapped ions, \href{https://www.nature.com/nature/journal/v476/n7359/full/nature10290.html}{Nature {\bf 476}, 181 (2011).}

\bibitem{Hanh19} H. Hahn, G. Zarantonello, M. Schulte, A. Bautista-Salvador, K. Hammerer, and C. Ospelkaus, Integrated $^9{\rm Be}^+$ multi-qubit gate device for the ion-trap quantum computer, \href{https://www.nature.com/articles/s41534-019-0184-5}{npj Quantum Information {\bf 5}, 70 (2019).}

\bibitem{Zarantonello19} G. Zarantonello, H. Hahn, J. Morgner, M. Schulte, A. Bautista-Salvador, R. F. Werner, K. Hammerer, and C. Ospelkaus, Robust and resource-efficient microwave near-field entangling $^9{\rm Be}^+$ gate, \href{https://journals.aps.org/prl/abstract/10.1103/PhysRevLett.123.260503}{Phys. Rev. Lett. {\bf 123}, 260503 (2019).}

\bibitem{Srinivas19} R. Srinivas, S. C. Burd, R. T. Sutherland, A. C. Wilson, D. J. Wineland, D. Leibfried, D. T. C. Allcock, and D. H. Slichter, Trapped-ion spin-motion coupling with microwaves and a near-motional oscillating magnetic field gradient, \href{https://journals.aps.org/prl/abstract/10.1103/PhysRevLett.122.163201}{Phys. Rev. Lett. {\bf122}, 163201 (2019).}

\bibitem{Sutherland19} R. T. Sutherland, R. Srinivas, S. C. Burd, D. Leibfried, A. C. Wilson, D. J. Wineland, D. T. C. Allcock, D. H. Slichter, and S. B. Libby, Versatile laser-free trapped-ion entangling gates, \href{https://iopscience.iop.org/article/10.1088/1367-2630/ab0be5}{New J. Phys. {\bf21}, 033033 (2019).}

\bibitem{Sutherland20} R. T. Sutherland, R. Srinivas, S. C. Burd, H. M. Knaack, A. C. Wilson, D. J. Wineland, D. Leibfried, D. T. C. Allcock, D. H. Slichter, and S. B. Libby, Laser-free trapped-ion entangling gates with simultaneous insensitivity to qubit and motional decoherence, \href{https://arxiv.org/abs/1910.14178}{arXiv.1910.14178}

\bibitem{JonathanPK00} D. Jonathan, M. B. Plenio and P. L. Knight, Fast quantum gates for cold trapped ions, \href{https://journals.aps.org/pra/abstract/10.1103/PhysRevA.62.042307}{Phys. Rev. A {\bf 62}, 042307 (2000).}

\bibitem{Szwer11} D. J. Szwer, S. C. Webster, A. M. Steane, and D. M. Lucas, Keeping a single qubit alive by experimental dynamic decoupling, \href{https://iopscience.iop.org/article/10.1088/0953-4075/44/2/025501}{J. Phys. B: At. Mol. Opt. Phys. {\bf44}, 25501 (2011).}

\bibitem{Piltz13} Ch. Piltz, B. Scharfenberger, A. Khromova, A.F. Var\'on, and Ch. Wunderlich, Protecting conditional quantum gates by robust dynamical decoupling, \href{https://journals.aps.org/prl/abstract/10.1103/PhysRevLett.110.200501}{Phys. Rev. Lett. {\bf 110}, 200501 (2013).}

\bibitem{Casanova15}
J. Casanova, Z.-Y. Wang, J. F. Haase, and M. B. Plenio, Robust dynamical decoupling sequences for individual-nuclear-spin addressing, \href{https://journals.aps.org/pra/abstract/10.1103/PhysRevA.92.042304}{Phys. Rev. A {\bf 92}, 042304 (2015).}

\bibitem{Puebla16} R. Puebla, J. Casanova, and M. B. Plenio, A robust scheme for the implementation of the quantum Rabi model in trapped ions, \href{http://iopscience.iop.org/article/10.1088/1367-2630/18/11/113039/meta;jsessionid=C9AFBB9E2406FECF8177DD4530814D5C.c3.iopscience.cld.iop.org}{New. J. Phys. {\bf 18}, 113039 (2016).}

\bibitem{Puebla17} R. Puebla, M.-J. Hwang, J. Casanova, and M. B. Plenio, Protected ultrastrong coupling regime of the two-photon quantum Rabi model with trapped ions, \href{https://journals.aps.org/pra/abstract/10.1103/PhysRevA.95.063844}{Phys. Rev. A {\bf 95}, 063844 (2017).}

\bibitem{Arrazola18} I. Arrazola, J. Casanova, J. S. Pedernales, Z.-Y. Wang, E. Solano, and M. B. Plenio, Pulsed dynamical decoupling for fast and robust two-qubit gates on trapped ions, \href{https://journals.aps.org/pra/cited-by/10.1103/PhysRevA.97.052312}{Phys. Rev. A {\bf97}, 052312 (2018).}

\bibitem{Wang19} Z.-Y. Wang, J. E. Lang, S. Schmitt, J. Lang, J. Casanova, L. McGuinness, T. S. Monteiro, F. Jelezko, and M. B. Plenio, Randomization of pulse phases for unambiguous and robust quantum sensing, \href{https://journals.aps.org/prl/abstract/10.1103/PhysRevLett.122.200403}{Phys. Rev. Lett. {\bf122}, 200403 (2019).}


\bibitem{Timoney11} N. Timoney, I. Baumgart, M. Johanning, A. F. Var\'on, M. B. Plenio, A. Retzker, and Ch. Wunderlich, Quantum gates and memory using microwave-dressed states, \href{https://www.nature.com/nature/journal/v476/n7359/full/nature10319.html}{Nature {\bf 476}, 185 (2011).}

\bibitem{Bermudez12} A. Bermudez, P. O. Schmidt, M. B. Plenio, and A. Retzker, Robust trapped-ion quantum logic gates by continuous dynamical decoupling, \href{https://journals.aps.org/pra/abstract/10.1103/PhysRevA.85.040302}{Phys. Rev. A {\bf 85}, 040302(R) (2012).}

\bibitem{Cohen15} I. Cohen, S. Weidt, W. K. Hensinger, and A. Retzker, Multi-qubit gate with trapped ions for microwave and laser-based implementation, \href{http://iopscience.iop.org/article/10.1088/1367-2630/17/4/043008/meta}{New. J. Phys. {\bf 17}, 043008 (2015).}

\bibitem{Wolk17} S. W\"olk and C. Wunderlich, Quantum dynamics of trapped ions in a dynamic field gradient using dressed states, \href{http://iopscience.iop.org/article/10.1088/1367-2630/aa7b22/meta}{New. J. Phys. {\bf 19}, 083021 (2017).}

\bibitem{Webb18} A. E. Webb, S. C. Webster, S. Collingbourne, D. Bretaud, A. M. Lawrence, S. Weidt, F. Mintert, and W. K. Hensinger, Resilient entangling gates for trapped ions, \href{https://journals.aps.org/prl/abstract/10.1103/PhysRevLett.121.180501}{Phys. Rev. Lett. {\bf121}, 180501 (2018).}

\bibitem{Weidt16} S. Weidt, J. Randall, S. C. Webster, K. Lake, A. E. Webb, I. Cohen, T. Navickas, B. Lekitsch, A. Retzker, and W.~K.~Hensinger, Trapped-ion quantum logic with global radiation fields, \href{https://journals.aps.org/prl/abstract/10.1103/PhysRevLett.117.220501}{Phys. Rev. Lett. {\bf 117}, 220501 (2016).}




\bibitem{Harty16} T. P. Harty, M. A. Sepiol, D. T. C. Allcock, C. J. Ballance, J. E. Tarlton, and D. M. Lucas, High-fidelity trapped-ion quantum logic using near-field microwaves, \href{https://journals.aps.org/prl/abstract/10.1103/PhysRevLett.117.140501}{Phys. Rev. Lett. {\bf 117}, 140501 (2016).}

\bibitem{CasanovaWS+18} J. Casanova, Z.-Y. Wang, I. Schwartz, and M.B. Plenio, Shaped pulses for energy-efficient high-field NMR at the nanoscale, \href{https://journals.aps.org/prapplied/abstract/10.1103/PhysRevApplied.10.044072}{Phys. Rev. Appl. {\bf 10}, 044072 (2018).}
 
\bibitem{CasanovaWS+19} J. Casanova, E. Torrontegui, M.B. Plenio, J.~J. Garc\'ia-Ripoll, and E. Solano, Modulated continuous wave control for energy-efficient electron-nuclear spin coupling, \href{https://journals.aps.org/prl/abstract/10.1103/PhysRevLett.122.010407}{Phys. Rev. Lett. {\bf 122}, 010407 (2019).}

\bibitem{Olmschenk07} S. Olmschenk, K. C. Younge, D. L. Moehring, D. N. Matsukevich, P. Maunz, and C. Monroe, Manipulation and detection of a trapped ${\rm Yb}^+$ hyperfine qubit, \href{https://journals.aps.org/pra/abstract/10.1103/PhysRevA.76.052314}{Phys. Rev. A {\bf 76}, 052314 (2007).}

\bibitem{Piltz14} C. Piltz, T. Sriarunothai, A.F. Var\'on, and C. Wunderlich, A trapped-ion-based quantum byte with $10^{-5}$ next-neighbour cross-talk, \href{https://www.nature.com/articles/ncomms5679}{Nat. Commun. {\bf5}, 4679 (2014).}


\bibitem{Roos08} C. F. Roos, Ion trap quantum gates with amplitude-modulated laser beams, \href{https://iopscience.iop.org/article/10.1088/1367-2630/10/1/013002}{New J. Phys. {\bf10}, 013002 (2008).}

\bibitem{Sorensen99} A. S\o rensen and K. M\o lmer, Quantum computation with ions in thermal motion, \href{https://journals.aps.org/prl/abstract/10.1103/PhysRevLett.82.1971}{Phys. Rev. Lett. {\bf 82}, 1971 (1999).}

\bibitem{Sorensen00} A. S\o rensen and K. M\o lmer, Entanglement and quantum computation with ions in thermal motion, \href{https://journals.aps.org/pra/abstract/10.1103/PhysRevA.62.022311}{Phys. Rev. A {\bf 62}, 022311 (2000).}

\bibitem{Solano99} E. Solano, R. L. de Matos Filho, and N. Zagury, Deterministic Bell states and measurement of the motional state of two trapped ions, \href{https://journals.aps.org/pra/abstract/10.1103/PhysRevA.59.R2539}{Phys. Rev. A {\bf59}, R2539(R) (1999).}


\bibitem{Brownnutt15} M. Brownnutt, M. Kumph, P. Rabl, and R. Blatt, Ion-trap measurements of electric-field noise near surfaces, \href{https://journals.aps.org/rmp/abstract/10.1103/RevModPhys.87.1419}{Rev. Mod. Phys. {\bf87}, 1419 (2015).}


\bibitem{Gillespie96} D. T. Gillespie, Exact numerical simulation of the Ornstein-Uhlenbeck process and its integral, \href{https://journals.aps.org/pre/abstract/10.1103/PhysRevE.54.2084}{Phys. Rev. E {\bf 54}, 2084 (1996).}

\bibitem{Cohen16} I. Cohen, N. Aharon, and A. Retzker, Continuous dynamical decoupling utilizing time-dependent detuning, \href{https://onlinelibrary.wiley.com/doi/abs/10.1002/prop.201600071}{Fortschr. Phys. {\bf 65}, 1600071 (2017).}









\end{thebibliography}
\end{document}